# The electromagnetic fields under, on and up Earth surface as earthquakes precursor in the Balkans and Black Sea regions

Strachimir Chterev Mavrodiev
Sofia, 14 October 2001
Institute for Nuclear Research and Nuclear Energy, Bulgarian Academy of Sciences
Blvd. Tzarigradsko shausee, 72, 1784 Sofia, Bulgaria
E-mail: Mavrodi@inrne.bas.bg

**Abstract**: It is demonstrate that the analysis of accuracy measurement of geomagnetic field and the behavior of local tide gravitational potential can serve as a earthquake precursor.

## Introduction

The prediction of the earthquakes'time (date, hour, minute), epicenter (region, depth), the amount of the emitted energy (Magnitude in Rihter scale), the destructive power on the surface and the danger from tsunami, has long and hard history. Very few are the successful predictions, when the scientists and the authorities have ventured to evacuate the risk areas, saving in this way thousands of lives [1].

Up till now the investigations of the earthquakes nature and the scientific possibilities for their prediction was a research subject for the geologists and geophysicists. In the last twentieth century in general the structure of the Earth, to a certain extend, the history of its formation and the lows of its development became clear. A model of the Earth's nucleus and its physical nature still does not exist, although a lot of its properties are known. Also there is not yet a correct model of the motion of the mantle, which causes the movements of the continental plaques. The data on the changes in the Earth's crust before, during and after an earthquake or vulcano's eruption are rather incorrect and uncompleted, both in respect to the space characteristics on the surface of the Earth crust and in its depth, as well as in time. The time dependence of this data is so stochastic (accidental), that it is very hard to deduce the useful information from the noise background, so the geological and geophysical data is not enough to make reliable predictions of the earthquakes [2].

There is no model for the origin and variations of the Earth's magnetic field, although it is studied in details on the surface of the Earth, as well, as over it due to the needs of air and ship navigation, the intercontinental telephone, radio and TV – communications and the systems of management and of the artificial satellites, the systems for strategic rocket defense [3]. Relatively well known are the daily, monthly (the usual tides) and yearly variations of the gravitational potential, which depend on the moment's positions of the bodies of the Solar system. [4].

The first steps in the investigations of the electric and magnetic properties by means of extra deep drilling (10 km into the few hundredths kilometers depth of the continental plaques) and electromagnetic measurements are already done, but for only a very small part of the Earth's crust [5]. The investigations of the electric currents on and under the Earth's surface are only local and sporadic. The electromagnetic properties of the ionosphere are well known theoretically and experimentally for the radio connections. The theory for the variations of the electrical potential in the heights of the atmosphere is well developed, but there is not detailed global data of its daily, monthly and yearly variations. From Franklin's time a lot is known, in the meteorology, about the exchange of the electrical charges between the Earth and its atmosphere through the thunders and lightning during the clouds forming and raining, but global data for the vertical currents in the atmosphere is still missing. The state of the radiation belt of the Earth in the near Cosmic space is well studied theoretically and also there is a global satellite monitoring of it, as a result the influence of the solar wind is taken into account and the changes of the Earth atmosphere, magnetic field and radiation belt, which follow from the eruptions of the Sun may be predicted. [6]. The influence on the meteorological processes of the wide atmospheric flows, generated by an elementary particle entering in to the atmosphere or by an atomic nucleus with high or extra high energy is not fully investigated.

**Methodology**

Thee building of "Nuclear winter" model, from which follows, that in a world's nuclear war the winners would not exist, because a death of the Earth's ecosystem will follow as result the end of today's human civilization, which is in reality the first important result of the birth of the new complex science-the environmental science.

This first success of the new science was the reason for the first precedent in history, when the leaders of the world following the recommendations and had agreed to stop the nuclear rearmament, for a future without arms, wars and violence. The model of the "Nuclear winter" was created thanks to the new computer and cosmic technologies, and of the impressive development of the physics, chemistry, biology in the sixties and seventies.

Here we have to mention the role of the physicist, chemist and biologist N. I. Vernadskiy [7], the theoretical physicist F. Dayson [8] and of the biologist N. M. Sisakyan [9] for the birth of the environmental science, whose subject is the whole Earth's ecosystem with its main component - the human civilization. Some people call this science "stable development". I would prefer the much more exact " harmonic existence" .

The efforts of an international group of scientists in the heighies to create scientific, managing, technological and business conditions for the harmonic existence of the Black sea's ecosystem were a real experience for a practical application of the new science [10].

**Hypothesis**

The hypothesis for possible correlation's between the earthquakes, the magnetic fields, the Earth's horizontal and vertical currents in the atmosphere was born when the historic data on the Black Sea was systemized. The fire pillars, observed during the Crimean (Кримското) earthquake in 1927, could be explained with the dehydration of the biogas condense at the time of the spread out of the earthquakes waves in the sediments and the ignition which follows in the presence of a high degree of ionization of the atmosphere. Such currents can be measured for example through a precise measurement of the Earth's magnetic fields. A common model for the origin of the attendant Earth's currents before, during and after an earthquake does not exist. Some scientists explain them with the piezoeffect, which generates local changes in the distribution of the free electric charges in the process when the intensities in the Earth's crust reach critical values, but there is no experimentally tested theoretical models of this process.

In December 1989 started, a continuos measurement of a projection of the Earth's magnetic field with a detector ( know- how of ОИЯИ, Дубна, Борис Василиев) with absolute precision less than 1 nano-Tesla at a sampling rate of 2.5 samples per second. Each minute the mean value (TMF), the mean value of the error (dTMF), the mean square deviation from the mean value (SigTMF) and the mean square deviation from the mean value of the error (SigdTMF) are evaluated, i.e. in 24 hours 1440 quartets of data are recorded. The choice of the recorded variables and the time for taking of their average values are obtained (determined) by the solution of the reverse problem for conforming the data for one projection of the magnetic field from our device with the standard data for the vector, measured in the geomagnetic station in Koprivshtiza of the Institute of Geophysics of BAS. The analysis of the data shows, that the big worlds and all the earthquakes close and strong enough can be "seen", but the interpretation is not unique. This means that they are recognized after we learn from the Internet that they had happened. The drop out of the quotation marks and the change of the verb recognized with predicted became possible after the seminar, which was held from the 23 till the 27 of July , 2001 in INRNE - BAS, Sofia [11]. The analysis of the magnetic data on the measurement of the Earth's currents in Volos, Greece and on the behavior of the tides gravitational potential on the Balkans, presented by Dr. K. Thanasoulas give us a possibility to predict in collaboration the day and the direction of earthquake near Island of Skiros . Only the direction of the earthquake was predicted, as Dr. K. Tanasoulas has only one device for measurement of the Earth's currents, but for the evaluation of the earthquakes' epicenter at least two of these are necessary.

The seminar proposed a Program for development and application of this method and when its usefulness is prooven its application in the Balkans and the region of Black Sea could be utilized. The choice of the experimental place is obviously the region of the Aegean Sea, because there two or tree earthquakes per day and between 10-20 earthquakes in a year occur with Richter's Magnitude bigger than 4. The Program requires at least three stationary devices for measurement of the magnetic field, the horizontal Earth's currents and the vertical currents in the atmosphere and one mobile for measurement of the magnetic field, and the vertical currents in the atmosphere in the region of the established future epicenter. If there are oportunities (financial and human resources) a monitoring of all possible other data on physical parameters, could be recommended. The data for the solar wind, the condition of the ionosphere, the radiation belt of the Earth, the local gravitational anomalies are needed in at least 3 km distances, the behavior of the water resources and their gas contains, the standard geological data for the intensities in the Earth's crust and few other are the obligatory for monitoring data. The data is stored in real time in the Center for collecting and analyzing it. The advanced modeling will answer the question how unique and statistically reliable are the predictions of the time, region and power of earthquake in two years. The preliminary estimations of the value of the hard ware of one facility is around 30000- 40000$, and of the whole complex around 250000$. There are no evaluations of the scientific input, but obviously the spectrum of the scientific qualifications should include theoretical physics, geophysics, geology, atmospheric sciences, geochemistry, mathematics, communication and computer technologies and also engineers for creation and exploitation of automated measuring systems with high relative and absolute precision, and velocity of the measurements up to 1000 Hz and high resistance for measurements of currents, electrical potentials, electrical resistance, magnetic fields and many other quantities.

The next picture is an example of Dr. Thanasoulas for determination of the direction towards a future epicenter of earthquake from the measurements of two orthogonal directions of the Earth's currents.

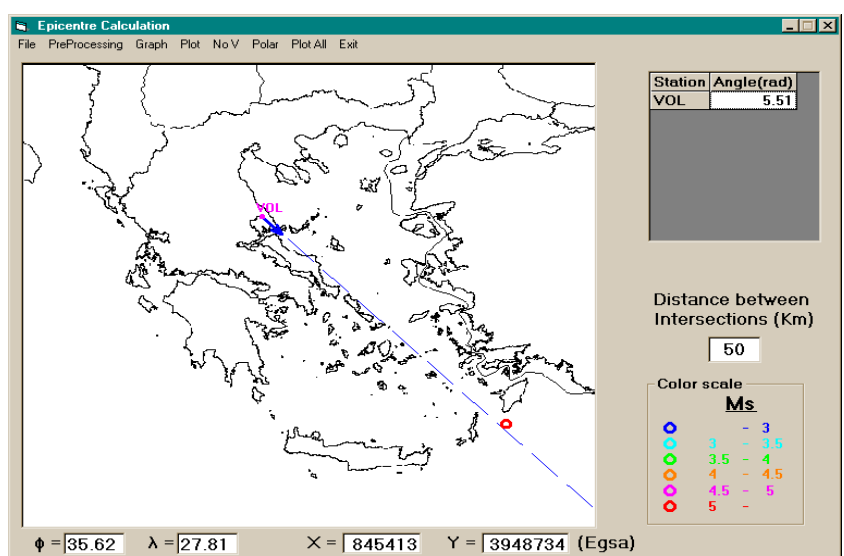

Figure 1.

On the following figure the behavior of the tide generating potential in the region of Yambol , Bulgaria , supplied by Prof. A. Venedikov from the Institute of Geophysics, BAS, Sofia [4].

Tide generating potential, Bulgaria, Yambol region
Hereafter is the figure for 01.01.2001 – 01.01.2002:

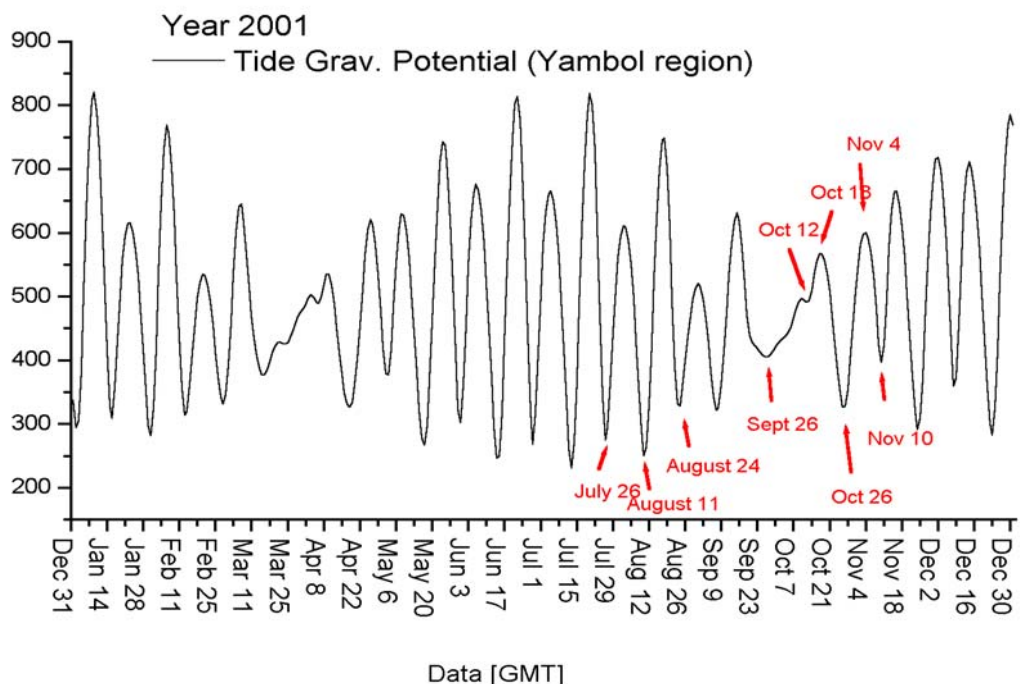

Figure 2.

**Results**

Figures 3-6 illustrate for the months July, August September and October, 2001, clearly and undoubtedly the correctness of the title. On each of them the 24 hours behavior of the measured projection of the magnetic field mean per minute is the up curve and its mean value per day is the dawn curve. On each Figure the dawn curve has a peak or peaks for a few days periods, which is a sign for a possible future earthquake or earthquakes in the region. The height of the peak (the value of the variation of the derivative) is directly proportional to the power and inverse proportional to the distance to the future epicenter. It depends on the geological structure (the isolines of the local gravitational anomalies) of the Earth's crust between the point of measurement and the epicenter. The possible date of the earthquake is determined from the next date, on which the Tide gravitational potential given at Figure 2 has a minimum (maximum) for the considered region. The probability for an earthquake is confirmed, if in the following few days for the 24 hour period, some specific variations or peaks appear of the magnetic field in contrast to its usual rather smooth behavior These signs are rather clearly seen on the 24 hours curves of the errors or the errors of the errors. On Fig.7 is presented a day with a normal 24 hours behavior of the magnetic field and its errors. On Fig.8 is given an example of 24 hour's curves, which is a confirmation of the prediction of the date of the event.

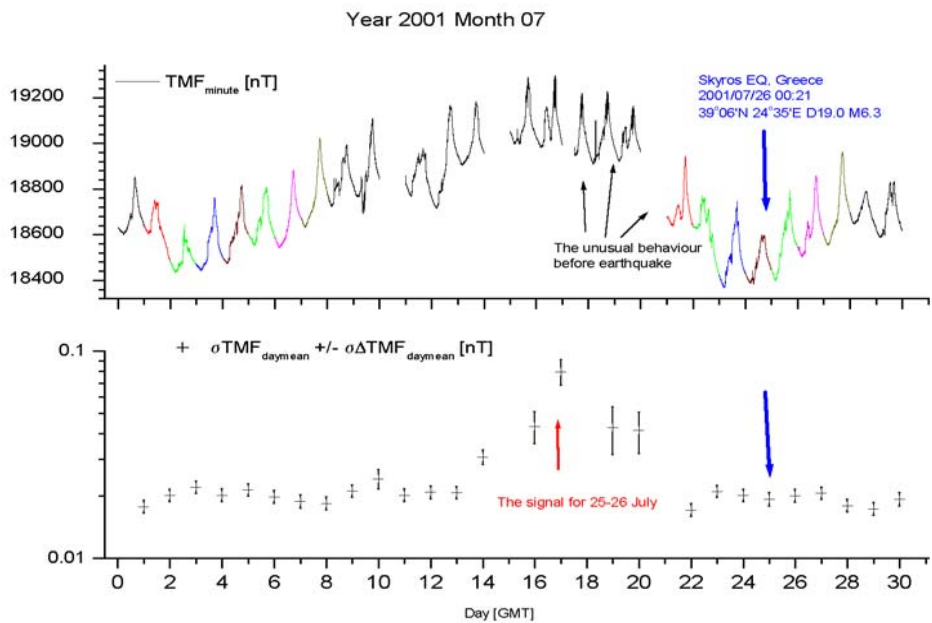

Figure 3.

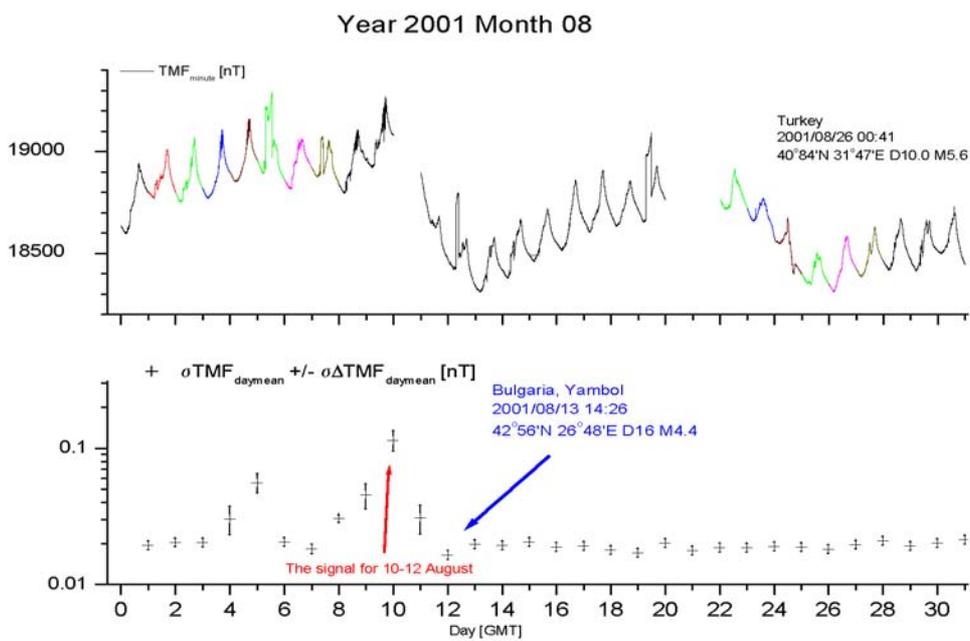

Figure 4.

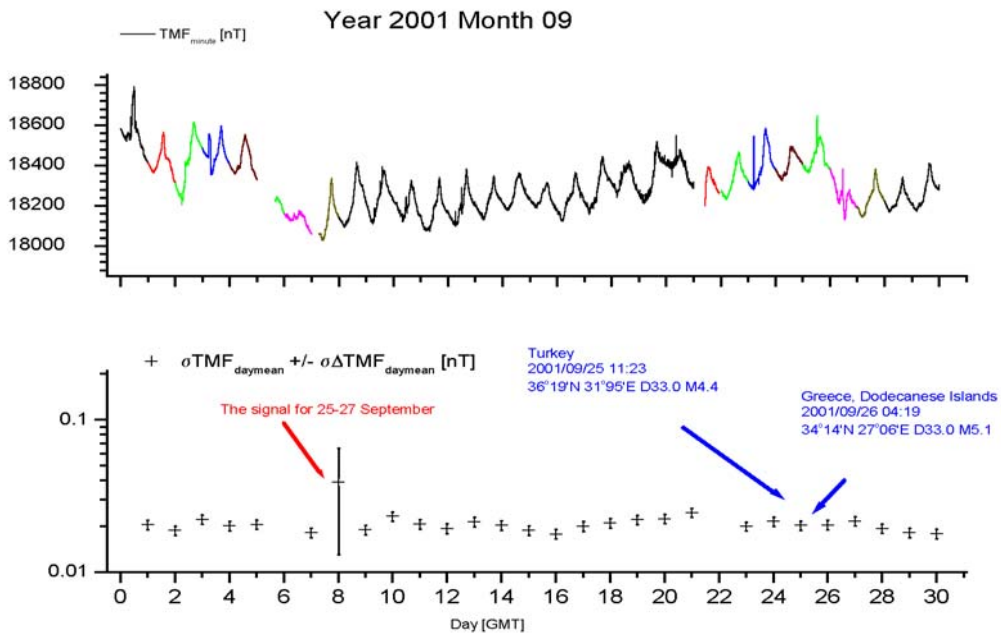

Figure 5.

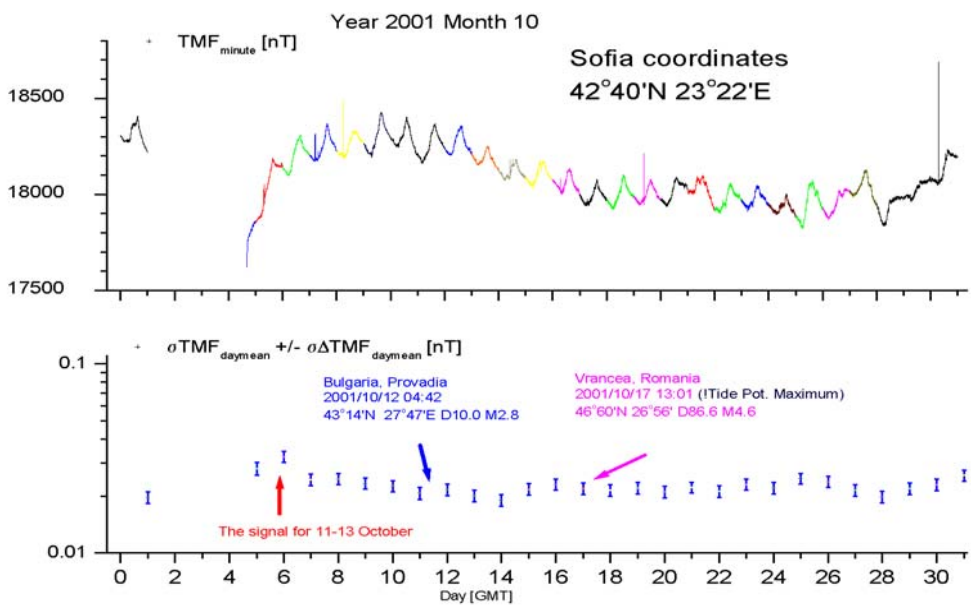

Figure 6.

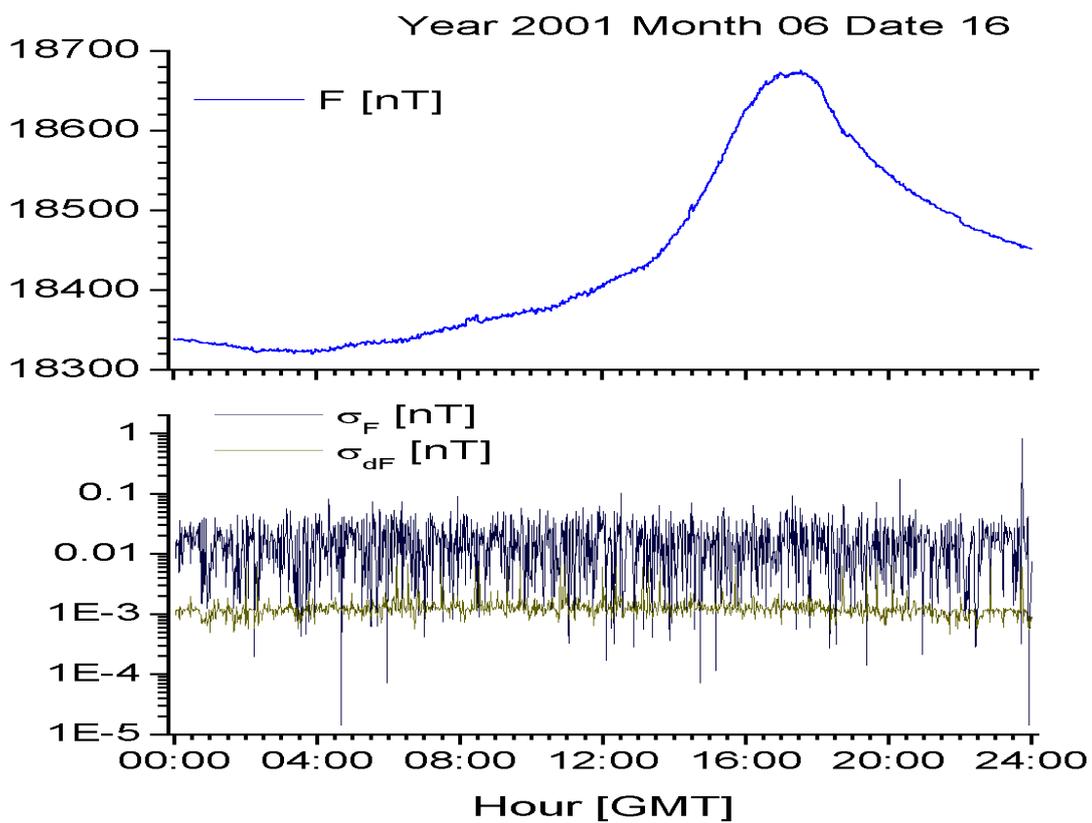

Figure 7,

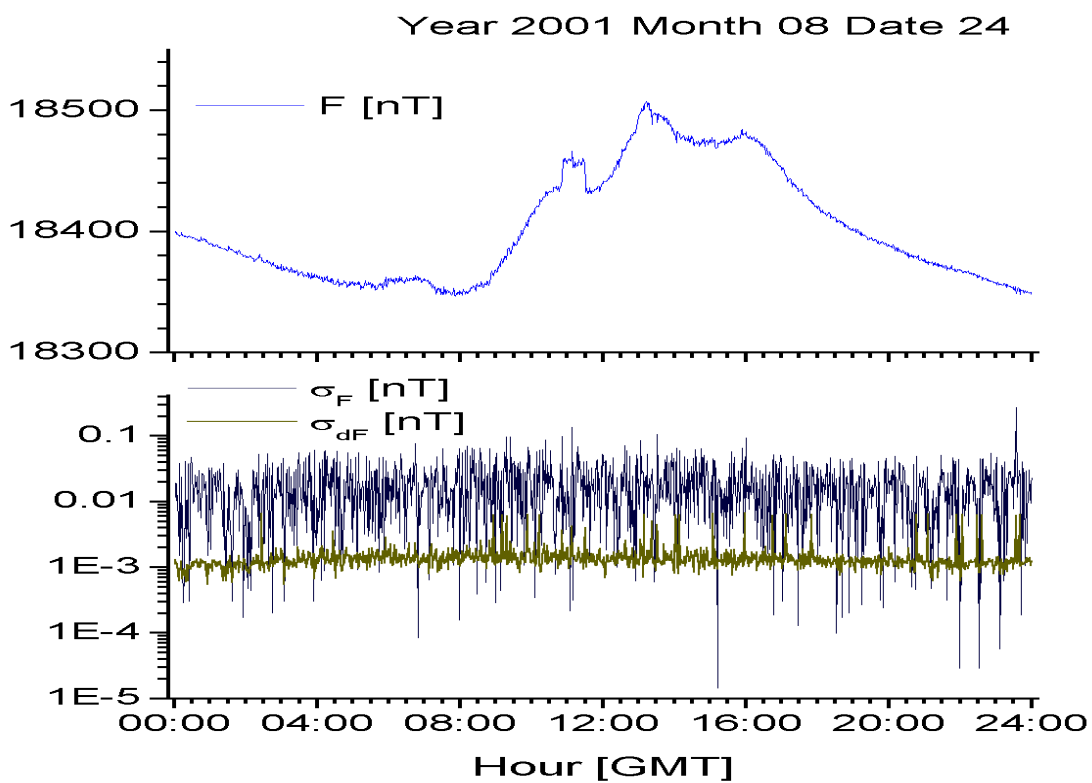

Figure 8.

The recently published works of scientists from the Moscow State Engineering and Ethnological Institute about a proved correlation between strong enough earth quakes and the change in of the density of the electrical charges in the radiation belt of the Earth (electrons and ionized atomic nuclei) [13], have turned into experimental fact our hypothesis about the physical correlations between the changes of the electromagnetic fields under, on and over the Earth's surface and the earthquakes. It looks like, that a quasi – northern (in the Northern Hemisphere) or quasi – southern (in the Southern Hemisphere) magnetic pole appears in the epicenter around the time of an earthquake. Until the end of 2001, Russia intents to send a special satellite for "prediction of earthquakes".

We would like that this work is accepted as a proven possibility of predicting the date of the earthquakes. On the Balkans each day a few earthquakes take place and until we are not able to give theoretical quantitative interpretation of the observed correlations and they are confirmed statistically, the question will stay open.

Why such a correlation in time, is not observed for the maximum of the tide's potential is explained by means of the greater probability for smoothing over the built up inhomogeneous in the tensor of tension at a lower density.

Why the weaker earthquakes in the region are only seen, but there is no any preliminary signal for them? A more clear answer of this question can be obtained by including in the monitoring the data about the three components of the magnetic field, the two components of the earth's currents, measured from two different points, the vertical electric current and the precise Fourier analysis together with the energetic evaluation of the contribution of the different part of the spectrum.

The theory must include the equations of the thermal, electromagnetic and gravitational state of the Earth and its atmosphere, the influence of the solar winds and the gravitational interactions in the Solar system. Even, the qualitative formulation of this complicated nonlinear, as a rule, system of equation looks like a dream for the future.

We would like this work to be considered as an additional argumentation of benefit of the organizing of the complex investigations and monitoring, proposed on the Seminar [10].

The Author is well aware of his limitations in expertise on the subject, resulting in the incompleteness of the references and will be very grateful to any one, who will help in filling up this omission. With pleasure I will accept all the opinions aiming at a certain degree unambiguous positive or negative answers of the considered problem.